\documentclass[aps,twocolumn,showpacs,preprintnumbers,amsmath,amssymb,
superscriptaddress]{revtex4}

\usepackage[T1]{fontenc}
\usepackage[latin9]{inputenc}
\usepackage{graphicx}

\begin{document}

\title{Rotating three-dimensional solitons in Bose Einstein condensates with gravity-like
attractive nonlocal interaction}
\author{F. Maucher}
\affiliation{Max-Planck-Institute for the Physics of Complex Systems, 01187 Dresden,
Germany}
\author{S. Skupin}
\affiliation{Max-Planck-Institute for the Physics of Complex Systems, 01187 Dresden,
Germany}
\affiliation{Friedrich-Schiller-University, Institute of Condensed Matter Theory and Solid State
Optics, 07743 Jena, Germany}
\author{M. Sheng}
\affiliation{Department of Physics, Shanghai University, Shanghai 200444, China}
\author{ W. Krolikowski}
\affiliation{Laser Physics Centre, Research School of Physics and Engineering,
Australian National University, Canberra, ACT 0200, Australia}

\begin{abstract}
We study formation of rotating three-dimensional high-order solitons (azimuthons) in Bose Einstein condensate  with
 attractive nonlocal nonlinear interaction. In particular, we demonstrate
formation
 of toroidal rotating solitons and investigate their stability. We show that
variational methods allow a very
  good approximation of such solutions and predict accurately the soliton rotation frequency.
   We also find that these rotating localized structures are  very robust and persist
  even if the initial condensate conditions are rather far from the exact soliton solutions.
  Furthermore, the presence of repulsive contact interaction does not prevent the existence of those
solutions, but allows to control their rotation.
We conjecture that self-trapped azimuthons are generic for condensates with attractive nonlocal interaction.
\end{abstract}
\maketitle

\section{Introduction\label{sec:intro}}
Studies of Bose Einstein condensates (BEC) belongs to one of the fastest developing research directions. The major
theoretical progress in this area has been stimulated by the fast experimental advances which enables to  investigate
subtle phenomena of fundamental nature~\cite{BEC_review},~\cite{BEC_review2}. In the semiclassical approach the spatial and temporal
evolution of the condensates wave function is commonly described by the Gross Pitaevskii equation~\cite{Dalfovo:99}
which reflects the interplay between kinetic energy of the condensate and the nonlinearity originating from the
interaction potential leading, among others, to the formation of localized structures, bright and dark
solitons~\cite{Khayakovich:science:02,Strecker:02:nature}. So far the main theoretical and experimental efforts have
been concentrating on condensates with contact (or hard-sphere) bosonic interaction which, in case of attraction,
may lead to collapse-like dynamics. Recently, also systems exhibiting a nonlocal, long-range dipolar
interaction~\cite{Goral:05} have attracted a significant attention. This interest has been stimulated by successful
condensation of Chromium atoms which exhibit an appreciable magnetic dipole
moment~\cite{Griesmaier:05,Beaufils:08:pra,Stuhler:05}. The presence of spatially nonlocal nonlinear
interaction and, at the same time, the ability to control
externally the character of local (contact) interactions via
the Feshbach resonance techniques offer the unique opportunity to study the effect of nonlocality on the dynamics,
stability and interaction of bright and dark matter wave
solitons~\cite{Pedri:05,Lahaye:07:nature,Koch:08:nphys,Pollack:09}.
The enhanced stability of localized
structures including fundamental, vortex and rotating solitons in nonlocal nonlinear media (not necessarily, BEC)
has been already pointed out in a number of theoretical
works~\cite{Bang:02:pre,Nath:07:pra,Nath:08,Cuevas:09:pra,Lashkin:07:pra,
Lashkin:08:pra_a,Zaliznyak:08:pla,Lashkin:09:pscr}.
In particular, stable toroidal solitons were presented
in~\cite{Zaliznyak:08:pla,Lashkin:09:pscr}.
 However, since the
dipole-dipole interaction is spatially
anisotropic, an additional trapping potential or a combination of attractive
two-particle and repulsive three-particle interaction were necessary.
Various trapping arrangements have been proposed to minimize or completely
eliminate this anisotropy. In
particular, O'Dell {\em at al.}~\cite{Odell:00} have recently suggested to use a series of triads of
orthogonally polarized laser beams illuminating cloud of cold atoms along three orthogonal axes so that the angular
dependence of the dipole-dipole nonlinear term is averaged out. The resulting nonlocal interaction potential becomes
effectively isotropic of the form $1/r$. It has been already shown by Turitsyn~\cite{Turitsyn:85:tmf} that a purely attractive
"gravitational" (or Coulomb) interaction potential prevents collapse of nonlinear localized waves and gives rise to
the formation of localized states - bright solitons which could be supported without necessity of using the external
trapping potential. If realized experimentally such trapping geometry would enable to study effects akin to
gravitational interaction. Few recent works have been dealing with this "gravitational" model of condensate looking,
among others, at the stability of localized structures such as fundamental solitons and two-dimensional
vortices~\cite{Giovanazzi:01:pra,Papadopoulos:07:pra,Cartarius:08:pra,Keles:08:pra}.

In this paper we study formation of three-dimensional high-order solitons in BEC with
gravity-like attractive nonlocal nonlinear potential. In particular, we demonstrate
formation of vortex tororidal solitons (solitons) and investigate their stability.
 We show that such BEC supports robust localized structures even if the initial conditions
  are rather far from the exact soliton solutions. Furthermore, we also demonstrate that
  the presence of repulsive    contact interaction does not prevent the existence of those
solutions, but allows to control their rotation.

The paper is organized as follows. In Sec.~\ref{model} we introduce briefly a
scaled nonlocal Gross-Pitaevskii equation
(GPE). We discuss two different response functions, the above long range $1/r$
response and the so-called Gaussian response
yielding a much shorter interaction range.
In Sec.~\ref{azimuthons} we recall general properties of rotating soliton
solutions (azimuthons), which are then approximated in
Sec.~\ref{variational} by means of a variational approach. Those variational approximations allow us to predict the rotation
frequency of the azimuthons which are then confronted with results from rigorous numerical simulations. Finally, self-trapped higher order
three-dimensional rotating solitons are presented in
Sec.~\ref{numerics}, and we show that such a nonlocal BEC's support robust
localized structures.

\section{Model \label{model}}

 We consider a Bose-Einstein atomic condensate with the  isotropic interatomic potential consisting of both,
repulsive contact as well as attractive long-range nonlocal interaction contributions. Following O'Dell~{\em
et.al}~\cite{Odell:00}, an attractive long-range interaction of ''gravitational`` form can be induced
by triads of frequency detuned laser beams resulting in the following dimensionless Gross-Pitaevskii
equation (GPE) for
the condensate wave function $\psi\left(\mathbf{r},t\right)$:
\begin{subequations}\label{eq:normalised_gpe}
\begin{align}
\partial_{t}\psi&=i\Delta\psi+i\Theta\psi\\
\Theta\left(\mathbf{r},t\right)&=\int\frac{\left|\psi\left(\mathbf{r}^{\prime},t\right)\right|^{2}}{\left|\mathbf{r}-
\mathbf{r}^{\prime}\right|}d^{3}r^{\prime}-\left|\psi\right|^{2}.
\end{align}
\end{subequations}
The nonlinear response $\Theta$ consists of both local and nonlocal contribution. Interestingly, for the
''gravitational`` nonlocal interaction $\Theta$  contains no additional parameter (see also Appendix~\ref{1/r}). The
ratio between local and nonlocal term is solely determined by the form of the wavefunction $\psi$. We will see later
(Sec.~\ref{variational}) that for very broad solitons the local contact interaction $\sim |\psi|^{2}$ becomes negligible.

In this paper we will also consider a second, different nonlocal model, the so-called Gaussian model of nonlocality.
Despite the fact that it is not motivated by a certain physical system, it serves as a popular toy model for the general
class of nonlocal Schr\"odinger (Gross-Pitaevskii) equations in one and two dimensional problems
\cite{Bang:02:pre,Krolikowski:04:job,Buccoliero:07,Buccoliero:07:PhysicaB,Skupin:08:oe}.
Here, we will extend this classical model to three transverse dimensions, and moreover allow an
additional local repulsive term similar to the previous case, and introduce
\begin{equation}
\label{eq:gauss}
\Theta\left(\mathbf{r},t\right)=\left(\frac {1}{2\pi}\right)^{3/2}\int
\left|\psi\left(\mathbf{r}^{\prime},t\right)\right|^{2} e^{-\frac
{\left|\mathbf{r}-
\mathbf{r}^{\prime}\right|^2}{2}}d^{3}r^{\prime}-\delta\left|\psi\right|^{2}.
\end{equation}
The additional parameter $\delta$ is necessary here to keep track of one of the two
degrees of freedom of the Gaussian response,
i.e. amplitude or width, which cannot be scaled out  (see Appendix \ref{normgauss}).
The value of $\delta$ determines the relative strength of the local repulsive term.
Note that compared to the above ''gravitational response, the interaction range of the Gaussian nonlocal response is
significantly shorter due to its rapid decay for $r \rightarrow \infty$.

As far as stability of localized states is concerned,
Turitsyn~\cite{Turitsyn:85:tmf} showed that the ground state of the nonlocal Schr{\"o}dinger equation with a
purely attractive $1/r$ kernel is stable (collapse arrest) using Lyapunoff's method.
A rather general estimate for non-negative responsefunctions has been found in \cite{ginibre:80:MZ} for arbitrary dimensions.
Bang {\em at al.}~\cite{Bang:02:pre} showed, using the same method,
that for systems with arbitrarily shaped, nonsingular response functions with positive definite Fourier spectrum, collapse cannot occur.
Obviously the stability of the ground state is only a necessary but not sufficient condition for the stability of
rotating higher-order states, which we will investigate in the following by means of numerical simulations.
In \cite{Froelich:02:CMP}, linear and global (modulational) stability under small perturbations of solutions
of the Hartree-equation was shown.

\section{Rotating solitons \label{azimuthons}}

It has been shown  earlier that azimuthons, i.e. multi-peak solitons with angular phase ramp
exhibit constant angular rotation  and hence can be represented
 by straightforward generalization of the usual (nonrotating) soliton  ansatz
 by including  an additional parameter, the angular frequency $\Omega$~\cite{Desyatnikov:05,Skryabin:07:pre}.
We write
\begin{equation}
\label{azimuthon_ansatz}
\psi (r,z,\phi ,t) = U(r,z,\phi -\Omega t)\mathrm{e}^{iE t},
\end{equation}
where $U$ is the complex amplitude  and $E$ is the  normalized chemical potential,
$r=\sqrt{x^{2}+y^{2}}$  and $\phi$ denotes the  azimuthal angle in the
plane  $(x,y)$.
It can be shown, that by inserting the above function  into the nonlocal GPE (\ref{eq:normalised_gpe})
one can derive  the formal relation
 for the rotation frequency~\cite{Rosanov:os:96:405,Skupin:08:oe}
\begin{equation}\label{omega}
\Omega=-\frac{IL-I^{\prime}M+XL-X^{\prime}M}{L^{2}-MM^{\prime}},
\end{equation}
where the functionals $M,M', X, X^{\prime},L,I, I'$ represent  the following integrals over the stationary
amplitude profiles of the azimuthons
\begin{subequations}\label{int_system}
\begin{align}
 M&=\int\left|U\right|^{2}\mathrm{d}^{3}\mathbf{r},\\
L&=-i\int U^{*}\partial_{\varphi}U\mathrm{d}^{3}\mathbf{r},\\
I&=\int U^{*}\Delta U\mathrm{d}^{3}\mathbf{r},\\
X&=\int \Theta \left( \mathbf {r} \right)
\left|U\left(\mathbf{r}\right)
\right|^{2}\mathrm{d}^{3}\mathbf{r},\\
M^{\prime}&=\int\left|\partial_{\varphi}U\right|^{2}\mathrm{d}^{3}\mathbf{r},\\
I^{\prime}&=i\int\partial_{\varphi}U^{*}\Delta U\mathrm{d}^{3}\mathbf{r},\\
X^{\prime}&=i\int \Theta \left( \mathbf {r} \right)
U\left(\partial_{\varphi}U^{*}\right)
\mathrm{d}^{3}\mathbf{r}.
\end{align}
\end{subequations}
The first two conserved functionals ($M$) and ($L$) have straightforward physical meanings of
 ''mass'' or ''number of particles'' and  ''angular momentum''.
In the next Section, we will compute approximate azimuthon solutions and their rotation frequency
employing  a certain ansatz for the stationary amplitude profile $U$.

\section{Variational approach \label{variational}}
In order to get some insight into possible localized states of the
Gross-Pitaevskii equation we resort first to the so called Lagrangian (or variational) approach~\cite{variational}.
It is easy to  show that  Eq.~(\ref{eq:normalised_gpe}) can be derived from the following Lagrangian
density:
\begin{equation}
\begin{split}
\mathcal{L}:=\frac{i}{2}\left(\psi\partial_{t}\psi^{*}-
\psi^{*}\partial_{t}\psi\right)+
\left|\nabla\psi\right|^{2}-
\frac{1}{2}\left|\psi\right|^{2}
\Theta \left( \mathbf {r} , t \right).
\end{split}
\label{lagr_density}
\end{equation}
It has been shown before that rotating solitons or 'azimuthons' are associated with nontrivial phase and amplitude
structure~\cite{Buccoliero:07}. In two-dimensional optical problems the simplest case represents the state falling between
optical vortex (ring-like pattern with $2\pi$ angular phase shift) and optical dipole in the form of two out-of-phase intensity
peaks~\cite{Buccoliero:07,Buccoliero:07:PhysicaB}.
In three dimensions, a reasonable ansatz for corresponding localized solutions is
\begin{equation}\label{ansatz2}
\begin{split}
\psi\left(r,z,\varphi,t\right) & := Ar\exp\left(-\frac{r^{2}+z^{2}}{2\sigma^{2}}\right)\mathrm{e}^{iEt} \\
& \quad \times \left[\cos\left(\varphi-\Omega t\right)+ip\sin\left(\varphi-\Omega t\right)\right],
\end{split}
\end{equation}
where parameter p varies between zero and unity. For $p=0$
 Eq.~(\ref{ansatz2}) describes a dipole structure consisting of two out-of-phase
 lobes, while for $p=1$ it is a three-dimensional vortex, i.e. toroid-like structure
 with zero in the center and azimuthal (in the $(x,y)$ plane) phase ramp of $2\pi$.
Using the ansatz Eq.~(\ref{ansatz2}), one can easily find that
\begin{equation}
IL-I^{\prime}M=0,
\end{equation}
which  shows that only the non-linear terms contribute to the frequency $\Omega$
 (vide formula Eq.~(\ref{omega})).

%

After inserting the  solution Eq.~(\ref{ansatz2})  into the Lagrangian density
$\mathcal{L}$,
and integrating over the whole 3D space we obtain  the Lagrangian $L$ which is the function
of variational parameters $\sigma$ and $A$ only. Looking for the extrema of $L$ leads to a set of
algebraic relations among the variational
variables.%

\subsection{The ''gravitational'' response}

\begin{figure}
\includegraphics[width=8cm]{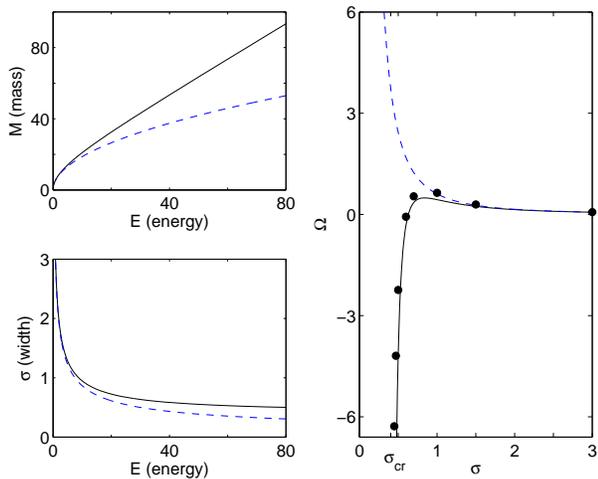}
\caption{(color online) The left panels show the dependency of the mass $M$ (top)
and the width $\sigma$ (bottom)
on the chemical potential $E$. Black curves show results from the variational approach including local repulsion,
dashed blue curves are without contact interaction. The right panel shows the angular frequency $\Omega$ as a function of $\sigma$.
Black dots denote results obtained from numerical simulations of the GPE (\ref{eq:normalised_gpe}). All plots are for $p=0.7$.}
\label{fig:properties_1_over_r}
\end{figure}

In this case, the amplitude $A$ can be expressed as a function of $p$ and $\sigma$
as follows (see also Appendix \ref{convolution})
 \begin{equation}\label{a_grav}
A^{2}=\frac{5\sqrt{2}\left(1+p^{2}\right)}{\frac{49p^{4}+86p^{2}+49}{120}\pi\sigma^{6}
-\frac{9p^{4}+6p^{2}+9}{32}\sigma^{4}},
\end{equation}
and the energy $E$ is given by
 \begin{equation}
E=
\frac{15\left[2 \pi \left(49 p^4+86p^2+49\right)-\frac{15 p^4-10
p^2-15}{2\sigma^2}\right]}
{4\sigma^2 \pi \left( 49 p^4+86 p^2+49 \sigma^2
\pi\right)-135-90p^2-135p^4}
\end{equation}
Because of the difference in the denominator,  the localized solution (with finite amplitude) exists only if
its width is greater than the critical value  $\sigma_{cr}\left(p\right)$,
\begin{equation}
\sigma_{cr}=\frac{3}{2}\sqrt{\frac{5}{\pi}}
\sqrt{\frac{3p^{4}+2p^{2}+3}{49p^{4}+86p^{2}+49}}.
\end{equation}
This threshold is an obvious consequence of competition between
nonlocal and  local interaction potentials, because the second term in the denominator of Eq.~(\ref{a_grav})
is due to the local contact interaction.
While the former being attractive, leads to spatial localization, the latter, which is repulsive,
tends to counteract it.  For small $\sigma$ the kinetic energy term is large and can be compensated only
if the particle density is high enough. In this regime the local repulsive
interaction prevails over the attraction leading to the expansion of the
condensate until the condition for its localization  (i.e. $\sigma > \sigma_{cr}$) is satisfied.
The rotation frequency  $\Omega$ is then given by the following relation
\begin{eqnarray}\label{eq:Omega}
\Omega &=& A^2\frac{\sigma^{2}p\sqrt{2}\left(4\sigma^{2}\pi-5\right)}{80}.
\end{eqnarray}
Interestingly, this expression is not sign definite, which means that we can expect both positive and negative rotation frequencies.
In particular, the azimuthon with the ''stationary'' width $\sigma_s=\sqrt{5/4\pi}\approx0.63$ has no angular velocity.
Again, this effect is due to competition between nonlocal and local contribution to $\Omega$ for small $\sigma$.
The nonlocal attractive interaction leads to a positive contribution to $\Omega$, the repulsive local interaction to a negative one.
The expression for $\Omega$ {\em without} repulsion can be obtained by the outlined variational procedure or by asymptotic expansion ($\sigma\rightarrow \infty$) up to
$\mathcal {O} \left( 1/\sigma^2 \right)$,
$\Omega  =  \frac{60(1+p^{2})p}{(49p^{4}+86p^{2}+49)\sigma^2}.$
As expected, this quantity is strictly positive. Both curves $\Omega$ versus $\sigma$ ($p=0.7$) with and without contact interaction
are shown in the right panel in Fig.~\ref{fig:properties_1_over_r}. We can see that the repulsive local interaction kicks in for $\sigma<1.5$.

\begin{figure}
\includegraphics[width=8cm]{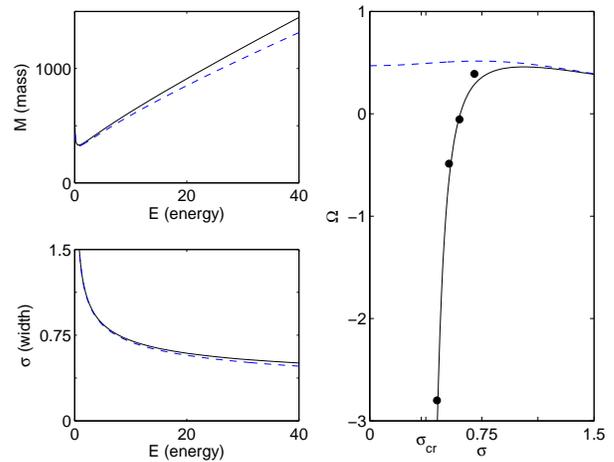}
\caption{(color online) Same as Fig.~\ref{fig:properties_1_over_r}, but for the Gaussian response given in Eq.~(\ref{eq:gauss}).
Black curves are for $\delta=0.01$, dashed blue curves without repulsion ($\delta=0.01$). All plots are for $p=0.7$.}
\label{fig:properties_gauss}
\end{figure}

As we observe in Fig.~\ref{fig:properties_1_over_r}, the mass behaves like $\sqrt{E}$ close to $E=0$
, since generally, $ M \sim A^2 \sigma^5 $, and for $\sigma \rightarrow \infty$, one finds $E\sim 1/\sigma^2 $ $M\sim1/\sigma $, whereas for $\sigma \rightarrow 0$, one finds $E\sim 1/\sigma^2 $, $M\sim\sigma$.
The fact that the mass can become zero for $E \rightarrow 0$ is a well-known property for very long range kernels,
 such as the Coulomb potential in three dimensions \cite{Froelich:02:CMP}.
For shorter ranged responses (e.g., Gaussian response, see Fig.~\ref{fig:properties_gauss}), the mass attains its minimum
at a finite value of $E$. In the limit of solely attractive local interaction ($E\rightarrow0$, $\sigma \rightarrow \infty$),
the mass is a monotonically decreasing function in $E$.

\subsection{The Gaussian response}

Repeating each step of the previous calculations for the Gaussian nonlocal response,
one ends up again with expressions for amplitude $A$ and rotation frequency $\Omega$,
given by
\begin{equation}
\label{eq:A_for_gauss_with_repulsion}
A^2=
\frac
{\sqrt{2}\left(1+p^2\right)\left(\sigma^2+1\right)^{9/2}}
{\frac{\left(9p^4+9+6p^2\right)\sigma^{13}}{160}
+\frac{\left(4p^2+1+p^4\right)\sigma^{11}}{20}
+\frac{\left(1+p^2\right)^2\sigma^9}{8}-\delta F_{\rm rep}}
\end{equation}
with $F_{\rm rep}=(\sigma^2+1)^{9/2}(9p^4+9+6p^2)\sigma^4/160$ and
\begin{equation}
\label{eq:omega_for_gauss_with_repulsion}
\Omega = A^2 \frac{p\left(\sigma^7-\delta(\sigma^2+1)^{7/2}\right)\sigma^2\sqrt{2}}{16(\sigma^2+1)^{7/2}}.
\end{equation}
As already pointed out in Sec.~\ref{model}, the additional parameter $\delta$ is necessary due to an additional degree of freedom
of the Gaussian response, and fixes the ratio between repulsion and attraction (see Appendix \ref{normgauss}).
Obviously, for $\delta=0$, the repulsive local contact interaction vanishes.
Here, $\sigma_{s} = \sqrt{\delta^{2/7} /\left(1-\delta^{2/7}\right)} \approx 0.60$ for $\delta=0.01$.

We observe that $E\sim1/\sigma^2$, $A\sim1/\sigma^2$, $M\sim\sigma\sim1/\sqrt{E}$ for both small and large $\sigma$.
Compared to the ''gravitational'' response, the range of this potential is much
shorter. Hence,
 when considering large $\sigma$ the Gaussian response acts more and more like a
local attractive response and higher order solitons become unstable (see end
of Sec.~\ref{numerics}).

\section{Numerical results \label{numerics}}

In this section, the predictions of the variational approach will be confronted with direct numerical simulations.
The approximate solitons resulting from the variational approach will be used as an initial conditions to our three-dimensional
code to compute their time evolution. In general, we find stable evolution, in particular the characteristic shape of the initial conditions is preserved.
For rotating azimuthons, the angular velocities will be measured and compared to the ones obtained in the previous section.

\begin{figure}
\includegraphics[width=8cm]{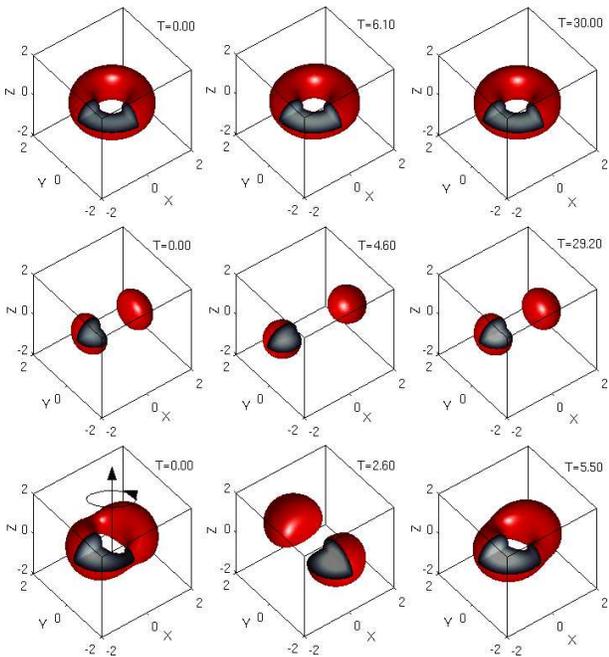}
\caption{(color online) Dynamics of the three-dimensional stable solitons in gravity-like BEC.
Iso-surfaces of the normalized density $\left|\psi\right|^{2}$ are depicted for different evolution times,
the interior density distribution is represented in grey-scales. The initial variational
parameters used are $\sigma=1$ and $p=1$ (torus, iso-density surface at $\left|\psi\right|^{2}=0.76$) for the upper row,
$p=0$ (dipole, iso-density surface at $\left|\psi\right|^{2}=1.41$)
for the middle one and finally $p=0.7$ (azimuthon, iso-density surface at $\left|\psi\right|^{2}=0.86$). The sense of the rotation ($\Omega=0.64$) is indicated by the arrows.}
\label{dynamics}
\end{figure}

In Fig.~\ref{dynamics} we illustrate the temporal evolution of three-dimensional solitons for the ''gravitational'' response, i.e.,
solutions to Eq.~(\ref{eq:normalised_gpe}). This first two rows present the classical stationary soliton solutions
torus and dipole, respectively. Due to imperfections of the initial conditions obtained from the variational approach we observe slight
oscillations upon evolution, in particular for the dipole solutions (second row). Those oscillations are not present if we use
numerically {\em exact} solutions (obtained from an iterative solver \cite{Skupin:pre:73:066603}) as initial conditions (not shown).
In the last row of Fig.~\ref{dynamics} we show the evolution of an azimuthon ($p=0.7, \sigma=1$). The rotation of the amplitude profile
is clearly visible. Again we observe radial oscillations due to the imperfect initial condition, but the solution is robust.

\begin{figure}
\includegraphics[width=8cm]{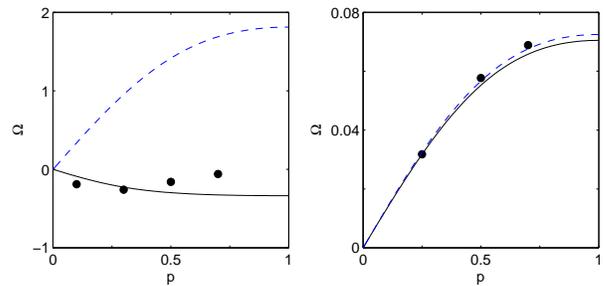}
\caption{(color online) Azimuthon rotation frequency $\Omega$ vs modulation parameter $p$
in gravity-like BEC, for $\sigma=0.6\approx\sigma_s$ (left panel) and $\sigma=3$ (right panel).
Black curves show results from the variational approach including local repulsion,
dashed blue curves are without contact interaction.
Black dots denote results obtained from numerical simulations of the GPE (\ref{eq:normalised_gpe}).
\label{fig:1_r_sigma_06}}
\end{figure}

\begin{figure}
\includegraphics[width=8cm]{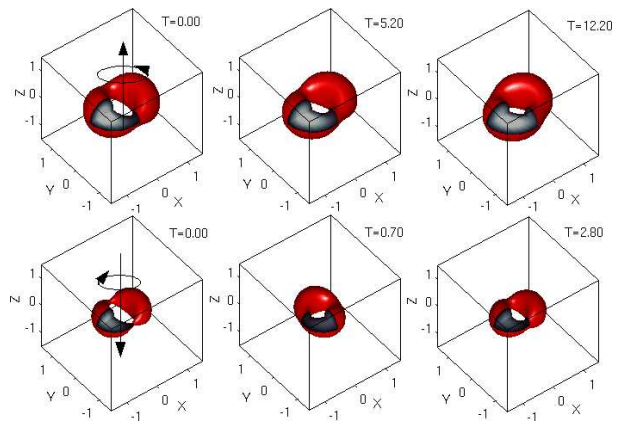}
\caption{(color online)
The upper row shows iso-density surfaces at $\left|\psi\right|^{2}=7.63$ for the very slow rotating
($\Omega\approx0$) azimuthon with
$p=0.7$ and $\sigma=0.61$. The lower row shows a fast counter-rotating
($\Omega=-2.24$) azimuthon with $p=0.7$, $\sigma=0.5$ and iso-density surface at $\left|\psi\right|^{2}=32$. Same plot style as in
Fig.~\ref{dynamics}.
\label{fig:1_r_standing}}
\end{figure}

Figure~\ref{fig:1_r_sigma_06} shows the dependency of the azimuthon rotation frequency as a function of the modulation
parameter $p$. Solid lines represent predictions from the variational model, black dots represent rotation frequency obtained
from numerical simulations. As expected from two-dimensional nonlocal models \cite{Lopez:ol:31:1100,Skupin:08:oe}, the modulus of $\Omega$ increases with $p$.
Our variational calculations predict that for small width $\sigma$, when repulsive interaction comes into play,
the sense of azimuthon rotation changes. In particular, we found a ''stationary'' width $\sigma_s$ where
the rotation frequency $\Omega$ vanishes. Indeed, full model simulations confirm this property,
since the first row in Fig.~\ref{fig:1_r_standing} and ~\ref{fig:1_r_sigma_06}, a) show a very slow rotation with opposite orientation,
so that the numerical stationary width is between $0.6\dots0.61$. Hence,
we propose that tuning
the strength of contact interaction in experiments allows to control the azimuthon rotation.

Furthermore, we observe that very narrow azimuthons ($\sigma \rightarrow \sigma_{cr}$)
have negative $\Omega$ and rotate very fast (see Fig.~\ref{fig:properties_1_over_r}).
This may be interesting for potential experiments, since the duration of BEC experiments
is restricted
to typically several hundreds of milliseconds.
 However, for azimuthons very close to
$\sigma_{cr}$ the ansatz function \ref{ansatz2} becomes less appropriate and using
variational initial conditions leads to very strong oscillations upon evolution, up to
the point where is is no longer possible to identify properly the rotation frequency $\Omega$.

\begin{figure}
\includegraphics[width=8cm]{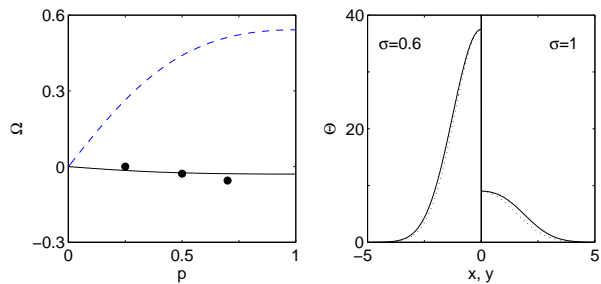}
\caption{(color online) Azimuthon rotation frequency $\Omega$ vs modulation
parameter $p$
in BEC with solely attractive Gaussian nonlocal response
for $\sigma=0.6\approx\sigma_s$ (left panel).
Black curves show results from the variational approach including local
repulsion ($\delta=0.01$),
dashed blue curves are without contact interaction ($\delta=0$).
Black dots denote results obtained from numerical simulations
(Eq.~\ref{eq:gauss}, $\delta=0.01$).
The right panel shows profiles of the convolution
term $\Theta$ [Eq.~(\ref{eq:gauss})] for $\sigma=0.6$ and $\sigma=1$.
Solid lines correspond to profiles along the major axis of the resulting
ellipsiod ($z=0$, $\varphi=0$), dotted lines to those along the minor axis
($z=0$, $\varphi=\pi/2$).
\label{fig:comparison_numerical_and_an_results_gaussian_response}}
\end{figure}

Concerning the Gaussian nonlocal response, we find very similar evolution
scenarios.
Results shown in
the left
panel in Fig.~\ref{fig:comparison_numerical_and_an_results_gaussian_response}
for the Gaussian response underline the observations from above, in particular
we also find nonrotating
azimuthons at $\sigma=\sigma_s$.
However, there are some important differences. First, it seems that our ansatz
Eq.~\ref{ansatz2}
is better suited for the Gaussian response, the radial oscillations we observed with
the ''gravitational'' are still present, but much weaker.
The second difference is due to the fact that the Gaussian response has a much shorter range
than the ''gravitational'' one.
For large $\sigma$ the Gaussian kernel acts like a attractive local response. As a consequence,
higher order solitons become unstable in the sense that the two humps spiral out. We observe unstable evolution
in numerical simulations for $\sigma\gtrsim0.9$ at $p=0.7$. The right panel in
Fig.~\ref{fig:comparison_numerical_and_an_results_gaussian_response} visualizes
the cause of this instability: For increasing sigma the resulting convolution
term $\Theta$ [Eq.~(\ref{eq:gauss})], which is responsible for the
self-trapping, becomes smaller in amplitude and asymmetric in the rotation
plane, which eventually leads to destabilization of the azimuthon.

\section{Conclusion}
We studied formation of rotating localized structures in Bose Einstein condensate
with different nonlocal interaction potentials.  We successfully used variational
techniques to investigate their dynamics and showed numerically that such
localized structures
are indeed robust objects which persist over long evolution times even if the initial
conditions significantly differ
from the exact soliton solutions.

For rotating solitons (azimuthons), we derived analytical expressions for the angular velocity,
in excellent agreement with rigorous three-dimensional numerical simulations.
Furthermore, we show that it is possible to control the rotation frequency by tuning
the local contact interaction, which is routinely possible by Feshbach resonance techniques.
In particular, we can change the sense of rotation, and we can find non-rotating
azimuthons.
We also identify parameter regions with particularly fast rotation, which may be
important for potential experimental observation of such solutions.

By using different nonlocal kernel functions we showed that rotating soliton solutions
are generic structures in nonlocal GPE's. Hence, we conjecture that the phenomena observed
in this paper are rather
universal and apply for a general class of attractive nonlocal interaction
potentials.

\appendix

\section{normalization of the ''gravitational'' model}\label{1/r}
We consider a Bose-Einstein atomic condensate with isotropic interatomic
potential consisting with both,
repulsive contact as well as attractive long-range nonlocal interaction contributions. Following O'Dell {\em
et.al}~\cite{Odell:00} the attractive long-range interaction which is electro-magnetically induced  by the triads of
frequency detuned laser beams with the intensity $I$ can be presented in the ''gravitational`` form
\begin{equation}
U_{Gr}\left(\tilde{\mathbf{r}}\right)=
-\frac{11}{4\pi}\frac{Iq^{2}\alpha^{2}}{c\epsilon_{0}}\frac{1}{\tilde{r}}=-\frac{u}{\tilde{r}}
\end{equation}
Here,  $q$ is the modulus of the wave vector,
$\alpha$ the isotropic, dynamic polarizability of the atoms, $c$
the light velocity and $\epsilon_{0}$ the permittivity of the free
space. Then the  complete  two-body interaction potential is given by
\begin{equation}
V\left(\tilde{\mathbf{r}}\right)=\frac{4\pi a\hbar^{2}}{m}\delta\left(\tilde{\mathbf{r}}\right)-\frac{u}{\tilde{r}},
\end{equation}
where the first term comes from the contact s-wave scattering,  $a$ is the
scattering length, and $m$ is the atomic mass. The potential
can only be written in this form, if the mean kinetic energy per particle
dominates the $u/\tilde{r}$-term, so that the short-range hard-sphere scattering
is not affected. This is fulfilled, if $a_{B}\ll\lambda_{B}\ll a$,
where $a_{B}:=\frac{h^{2}}{mu}$ is the Bohr radius, associated with
the interaction, and $\lambda_{B}$ is the de Broglie wavelength.

The temporal and spatial dynamics of the condensate wave function
$\tilde{\psi}\left(\mathbf{\tilde{r}},\tilde{t}\right)$
is then governed by the the following Gross-Pitaevskii equation (GPE):
\begin{equation}
\begin{split}
\!\!i\hbar\partial_{\tilde{t}}\tilde{\psi}+
\frac{\hbar^{2}}{2m}\Delta_{\tilde{\mathbf{r}}}\tilde{\psi}
+u\int\frac{\left|\tilde{\psi}\left(\tilde{\mathbf{r}}^{\prime},\tilde{t}\right)\right|^{2}}{
\left|\tilde{\mathbf{r}}-
\tilde{\mathbf{r}}^{\prime}\right|}d^{3}
\tilde{\mathbf{r}}^{\prime}\tilde{\psi} & \\
- \frac{4\pi
a\hbar^{2}}{m}\left|\tilde{\psi}\right|^{2}
\tilde{\psi} &=0.
\end{split}
\end{equation}
Using the normalization
\begin{subequations}
\begin{align}
\psi&:=\sqrt{8\pi aR_{c}^{2}}\tilde{\psi}=:\frac{1}{\psi_{c}}\tilde{\psi} \\
\mathbf{r}&:=\sqrt{\frac{um}{4\pi
a\hbar^{2}}}\tilde{\mathbf{r}}=:\frac{1}{R_{c}}\tilde{\mathbf{r}} \\
t&:=\frac{\hbar}{2mR_{c}^{2}}\tilde{t}=:\frac{1}{T_{c}}\tilde{t},
\end{align}
\end{subequations}
one ends up with the dimensionless GPE (\ref{eq:normalised_gpe}).
The actual values for the scaling parameters are $R_{c}=19\mu m$,
$m=3.8\cdot10^{-26}\mathrm{kg}$, $a=3nm$, $u=2\cdot10^{-13}eVnm$,
$T{}_{c}=0.25s$ which corresponds to typical experimental conditions~\cite{Odell:00}.
Then one finds that for the condensate consisting of, say, $N=10000$ atoms
the above normalization gives
\begin{equation}
M=\int\left|\psi\right|^{2}\mathrm{d}^{3}\mathbf{r}=\frac{N}{\psi_{c}^{2}R_{c}^{
3 }}\approx{42}.
\end{equation}

\section{normalization for the Gaussian model}\label{normgauss}
For the Gaussian response, one has to start from the equation
\begin{equation}
\begin{split}
i\hbar\partial_{\tilde{t}}\tilde{\psi}+
\frac{\hbar^{2}}{2m}\Delta\tilde{\psi}
-\frac{4\pi
a\hbar^{2}}{m}\left|\tilde{\psi}\right|^{2}\tilde{\psi} & \\
+A_{R}\int
\left|\tilde{\psi}\left(\tilde{\mathbf{r}}^{\prime},\tilde{t}\right)\right|^{2}
e^{-\frac{\left| \tilde{\mathbf{r}}-\tilde{\mathbf{r}}^{\prime} \right|^2}{2\sigma_{R}^2}}
\mathrm{d}^{3} \tilde{\mathbf{r}}^{\prime}\tilde{\psi} & =0
\end{split}
\end{equation}
Here, the degree of nonlocality (i.e. $\sigma_{R}$), that is fixed for the $1/r$-response, and the amplitude $A_R$ of the the response function can be chosen.
By using the normalization
\begin{subequations}
\begin{align}
\psi&:=
\frac{1}{4}\frac{\hbar 2^{3/4}}{\sigma^{2}\pi^{3/4}\sqrt{A_R\sigma m}}
=: \frac{1}{\psi_c}\tilde{\psi} \\
\mathbf{r}&:= \sigma_{R} \tilde{\mathbf{r}} =: \frac{1}{R_c}\tilde{\mathbf{r}} \\
t&:= \frac{2\sigma^{2}m}{\hbar}\tilde{t}
=:\frac{1}{T_{c}}\tilde{t} \\
\delta & := \frac{a\hbar^{2}\sqrt{2}}{\sigma_{R}^3\sqrt{\pi}mA_{R}},
\end{align}
\end{subequations}
one ends up with Eq.~\ref{eq:gauss},
that has the additional degree of freedom $\delta$.

\section{convolution}\label{convolution}
The convolution term in the Lagrangian (\ref{lagr_density}) can be calculated
analytically by, for instance, re-writing the
integrand in terms of spherical harmonics  $Y_{ij}$.
%
%
%
%
Then,
$\left|\psi\left(\mathbf{r}\right)\right|^{2}=\sum y_{ij}Y_{ij}$,
where $y_{ij}$ denote the coefficients of the spherical harmonics, and
the convolution integral can be easily calculated leading to the
following
result:
\begin{align*}
& \frac{1}{2}\int\left|\psi\left(\mathbf{r}\right)\right|^{2}
\int\frac{\left|\psi\left(\mathbf{r}^{\prime}\right)\right|^{2}}{\left|\mathbf{r
}- \mathbf{r}^{\prime}\right|}d^{3}\mathbf{r}^{\prime}d^{3}\mathbf{r}\\
= &
\frac{1}{2}\int_{0}^{\infty}\int_{0}^{r} 4\pi r A^{4}r^{\prime2}
\exp\left(-
\frac{r^{\prime2}+r^{2}}{\sigma^{2}}\right) \\
 &
\times\left(y_{00}^{2}+\frac{1}{5}y_{20}^{2}\left(\frac{r^{\prime}}{r}\right)^{2
}
+\frac{2}{5}y_{2\pm2}^{2}\left(\frac{r^{\prime}}{r}\right)^{2}\right)r^{\prime2}
\mathrm{d}r^{\prime}r^{2}\mathrm{d}r\\
 &
+\frac{1}{2}\int_{0}^{\infty}\int_{r}^{\infty}4\pi r^{\prime} A^{4} r^2 \exp\left(-\frac{
r^{\prime2}+r^2}
 {\sigma^{2}}\right) \\
 &
 \times
\left(y_{00}^{2}+\frac{1}{5}y_{20}^{2}\left(\frac{r}{r^{\prime}}\right)^{ 2}
 +\frac{2}{5}y_{2\pm2}^{2}\left(\frac{r}{r^{\prime}}\right)^{2}\right)r^{\prime2
}\mathrm{d}r^{\prime}r^{2}\mathrm{d}r\\
 = & \frac{49+86p^{2}+49p^{4}}{240}
\sigma^{9}\pi^{\frac{5}{2}}A^{4}\frac{1}{\sqrt{2}}.
\label{eq:L4}
\end{align*}

\bibliography{BEC}

\begin{thebibliography}{39}
\expandafter\ifx\csname natexlab\endcsname\relax\def\natexlab#1{#1}\fi
\expandafter\ifx\csname bibnamefont\endcsname\relax
  \def\bibnamefont#1{#1}\fi
\expandafter\ifx\csname bibfnamefont\endcsname\relax
  \def\bibfnamefont#1{#1}\fi
\expandafter\ifx\csname citenamefont\endcsname\relax
  \def\citenamefont#1{#1}\fi
\expandafter\ifx\csname url\endcsname\relax
  \def\url#1{\texttt{#1}}\fi
\expandafter\ifx\csname urlprefix\endcsname\relax\def\urlprefix{URL }\fi
\providecommand{\bibinfo}[2]{#2}
\providecommand{\eprint}[2][]{\url{#2}}

\bibitem[{\citenamefont{Giorgini et~al.}(2008)\citenamefont{Giorgini,
  Pitaevskii, and Stringari}}]{BEC_review}
\bibinfo{author}{\bibfnamefont{S.}~\bibnamefont{Giorgini}},
  \bibinfo{author}{\bibfnamefont{L.~P.} \bibnamefont{Pitaevskii}},
  \bibnamefont{and}
  \bibinfo{author}{\bibfnamefont{S.}~\bibnamefont{Stringari}},
  \bibinfo{journal}{Rev. Mod. Phys.} \textbf{\bibinfo{volume}{80}},
  \bibinfo{pages}{1215} (\bibinfo{year}{2008}).

\bibitem[{\citenamefont{Bloch et~al.}(2008)\citenamefont{Bloch, Dalibard, and
  Zwerger}}]{BEC_review2}
\bibinfo{author}{\bibfnamefont{I.}~\bibnamefont{Bloch}},
  \bibinfo{author}{\bibfnamefont{J.}~\bibnamefont{Dalibard}}, \bibnamefont{and}
  \bibinfo{author}{\bibfnamefont{W.}~\bibnamefont{Zwerger}},
  \bibinfo{journal}{Rev. Mod. Phys.} \textbf{\bibinfo{volume}{80}},
  \bibinfo{pages}{885} (\bibinfo{year}{2008}).

\bibitem[{\citenamefont{Dalfovo et~al.}(1999)\citenamefont{Dalfovo, Giorgini,
  Pitaevski, and Stringari}}]{Dalfovo:99}
\bibinfo{author}{\bibfnamefont{F.}~\bibnamefont{Dalfovo}},
  \bibinfo{author}{\bibfnamefont{S.}~\bibnamefont{Giorgini}},
  \bibinfo{author}{\bibfnamefont{L.~P.} \bibnamefont{Pitaevski}},
  \bibnamefont{and}
  \bibinfo{author}{\bibfnamefont{S.}~\bibnamefont{Stringari}},
  \bibinfo{journal}{Rev. Mod. Phys.} \textbf{\bibinfo{volume}{71}},
  \bibinfo{pages}{463} (\bibinfo{year}{1999}).

\bibitem[{\citenamefont{Khaykovich et~al.}(2002)\citenamefont{Khaykovich,
  Schreck, Ferrari, Bourdel, Cubizolles, Carr, Castin, and
  Salomon}}]{Khayakovich:science:02}
\bibinfo{author}{\bibfnamefont{L.}~\bibnamefont{Khaykovich}},
  \bibinfo{author}{\bibfnamefont{F.}~\bibnamefont{Schreck}},
  \bibinfo{author}{\bibfnamefont{G.}~\bibnamefont{Ferrari}},
  \bibinfo{author}{\bibfnamefont{T.}~\bibnamefont{Bourdel}},
  \bibinfo{author}{\bibfnamefont{J.}~\bibnamefont{Cubizolles}},
  \bibinfo{author}{\bibfnamefont{L.~D.} \bibnamefont{Carr}},
  \bibinfo{author}{\bibfnamefont{Y.}~\bibnamefont{Castin}}, \bibnamefont{and}
  \bibinfo{author}{\bibfnamefont{C.}~\bibnamefont{Salomon}},
  \bibinfo{journal}{Science} \textbf{\bibinfo{volume}{296}},
  \bibinfo{pages}{1290} (\bibinfo{year}{2002}).

\bibitem[{\citenamefont{Strecker et~al.}(2002)\citenamefont{Strecker,
  Partridge, Truscott, and Hulet}}]{Strecker:02:nature}
\bibinfo{author}{\bibfnamefont{K.~E.} \bibnamefont{Strecker}},
  \bibinfo{author}{\bibfnamefont{G.~B.} \bibnamefont{Partridge}},
  \bibinfo{author}{\bibfnamefont{A.~G.} \bibnamefont{Truscott}},
  \bibnamefont{and} \bibinfo{author}{\bibfnamefont{R.~G.} \bibnamefont{Hulet}},
  \bibinfo{journal}{Nature London} \textbf{\bibinfo{volume}{417}},
  \bibinfo{pages}{150} (\bibinfo{year}{2002}).

\bibitem[{\citenamefont{Goral et~al.}(2000)\citenamefont{Goral, Rzazewski, and
  Pfau}}]{Goral:05}
\bibinfo{author}{\bibfnamefont{K.}~\bibnamefont{Goral}},
  \bibinfo{author}{\bibfnamefont{K.}~\bibnamefont{Rzazewski}},
  \bibnamefont{and} \bibinfo{author}{\bibfnamefont{T.}~\bibnamefont{Pfau}},
  \bibinfo{journal}{Phys. Rev. A} \textbf{\bibinfo{volume}{61}},
  \bibinfo{pages}{051601(R)} (\bibinfo{year}{2000}).

\bibitem[{\citenamefont{Beaufils et~al.}(2008)\citenamefont{Beaufils,
  Chicireanu, Zanon, Laburthe-Tolra, Mar\'{e}chal, Vernac, Keller, and
  Gorceix}}]{Beaufils:08:pra}
\bibinfo{author}{\bibfnamefont{Q.}~\bibnamefont{Beaufils}},
  \bibinfo{author}{\bibfnamefont{R.}~\bibnamefont{Chicireanu}},
  \bibinfo{author}{\bibfnamefont{T.}~\bibnamefont{Zanon}},
  \bibinfo{author}{\bibfnamefont{B.}~\bibnamefont{Laburthe-Tolra}},
  \bibinfo{author}{\bibfnamefont{E.}~\bibnamefont{Mar\'{e}chal}},
  \bibinfo{author}{\bibfnamefont{L.}~\bibnamefont{Vernac}},
  \bibinfo{author}{\bibfnamefont{J.-C.} \bibnamefont{Keller}},
  \bibnamefont{and} \bibinfo{author}{\bibfnamefont{O.}~\bibnamefont{Gorceix}},
  \bibinfo{journal}{Phys. Rev. A} \textbf{\bibinfo{volume}{77}},
  \bibinfo{pages}{061601(R)} (\bibinfo{year}{2008}).

\bibitem[{\citenamefont{Griesmaier et~al.}(2005)\citenamefont{Griesmaier,
  Werner, Hensler, Stuhler, and Pfau}}]{Griesmaier:05}
\bibinfo{author}{\bibfnamefont{A.}~\bibnamefont{Griesmaier}},
  \bibinfo{author}{\bibfnamefont{J.}~\bibnamefont{Werner}},
  \bibinfo{author}{\bibfnamefont{S.}~\bibnamefont{Hensler}},
  \bibinfo{author}{\bibfnamefont{J.}~\bibnamefont{Stuhler}}, \bibnamefont{and}
  \bibinfo{author}{\bibfnamefont{T.}~\bibnamefont{Pfau}},
  \bibinfo{journal}{Phys. Rev. Lett.} \textbf{\bibinfo{volume}{94}},
  \bibinfo{pages}{160401} (\bibinfo{year}{2005}).

\bibitem[{\citenamefont{Stuhler et~al.}(2005)\citenamefont{Stuhler, Griesmaier,
  Koch, Fattori, Pfau, Giovanazzi, Pedri, and Santos}}]{Stuhler:05}
\bibinfo{author}{\bibfnamefont{J.}~\bibnamefont{Stuhler}},
  \bibinfo{author}{\bibfnamefont{A.}~\bibnamefont{Griesmaier}},
  \bibinfo{author}{\bibfnamefont{T.}~\bibnamefont{Koch}},
  \bibinfo{author}{\bibfnamefont{M.}~\bibnamefont{Fattori}},
  \bibinfo{author}{\bibfnamefont{T.}~\bibnamefont{Pfau}},
  \bibinfo{author}{\bibfnamefont{S.}~\bibnamefont{Giovanazzi}},
  \bibinfo{author}{\bibfnamefont{P.}~\bibnamefont{Pedri}}, \bibnamefont{and}
  \bibinfo{author}{\bibfnamefont{L.}~\bibnamefont{Santos}},
  \bibinfo{journal}{Phys. Rev. Lett.} \textbf{\bibinfo{volume}{95}},
  \bibinfo{pages}{150406} (\bibinfo{year}{2005}).

\bibitem[{\citenamefont{Lahaye et~al.}(2007)\citenamefont{Lahaye, Koch,
  Fr{\"o}hlich, Fattori, Metz, Griesmaier, Giovanazzi, and
  Pfau}}]{Lahaye:07:nature}
\bibinfo{author}{\bibfnamefont{T.}~\bibnamefont{Lahaye}},
  \bibinfo{author}{\bibfnamefont{T.}~\bibnamefont{Koch}},
  \bibinfo{author}{\bibfnamefont{B.}~\bibnamefont{Fr{\"o}hlich}},
  \bibinfo{author}{\bibfnamefont{M.}~\bibnamefont{Fattori}},
  \bibinfo{author}{\bibfnamefont{J.}~\bibnamefont{Metz}},
  \bibinfo{author}{\bibfnamefont{A.}~\bibnamefont{Griesmaier}},
  \bibinfo{author}{\bibfnamefont{S.}~\bibnamefont{Giovanazzi}},
  \bibnamefont{and} \bibinfo{author}{\bibfnamefont{T.}~\bibnamefont{Pfau}},
  \bibinfo{journal}{Nature (London)} \textbf{\bibinfo{volume}{448}},
  \bibinfo{pages}{672} (\bibinfo{year}{2007}).

\bibitem[{\citenamefont{Koch et~al.}(2008)\citenamefont{Koch, Lahaye, Metz,
  Fr{\"o}hlich, Griesmaier, and Pfau}}]{Koch:08:nphys}
\bibinfo{author}{\bibfnamefont{T.}~\bibnamefont{Koch}},
  \bibinfo{author}{\bibfnamefont{T.}~\bibnamefont{Lahaye}},
  \bibinfo{author}{\bibfnamefont{J.}~\bibnamefont{Metz}},
  \bibinfo{author}{\bibfnamefont{B.}~\bibnamefont{Fr{\"o}hlich}},
  \bibinfo{author}{\bibfnamefont{A.}~\bibnamefont{Griesmaier}},
  \bibnamefont{and} \bibinfo{author}{\bibfnamefont{T.}~\bibnamefont{Pfau}},
  \bibinfo{journal}{Nat. Phys.} \textbf{\bibinfo{volume}{4}},
  \bibinfo{pages}{218} (\bibinfo{year}{2008}).

\bibitem[{\citenamefont{Pedri and Santos}(2005)}]{Pedri:05}
\bibinfo{author}{\bibfnamefont{P.}~\bibnamefont{Pedri}} \bibnamefont{and}
  \bibinfo{author}{\bibfnamefont{L.}~\bibnamefont{Santos}},
  \bibinfo{journal}{Phys. Rev. Lett.} \textbf{\bibinfo{volume}{95}},
  \bibinfo{pages}{200404} (\bibinfo{year}{2005}).

\bibitem[{\citenamefont{Pollack et~al.}(2009)\citenamefont{Pollack, Dries,
  Junker, Chen, Corcovilos, and Hulet}}]{Pollack:09}
\bibinfo{author}{\bibfnamefont{S.~E.} \bibnamefont{Pollack}},
  \bibinfo{author}{\bibfnamefont{D.}~\bibnamefont{Dries}},
  \bibinfo{author}{\bibfnamefont{M.}~\bibnamefont{Junker}},
  \bibinfo{author}{\bibfnamefont{Y.~P.} \bibnamefont{Chen}},
  \bibinfo{author}{\bibfnamefont{T.~A.} \bibnamefont{Corcovilos}},
  \bibnamefont{and} \bibinfo{author}{\bibfnamefont{R.~G.} \bibnamefont{Hulet}},
  \bibinfo{journal}{Phys. Rev. Lett.} \textbf{\bibinfo{volume}{102}},
  \bibinfo{pages}{090402} (\bibinfo{year}{2009}).

\bibitem[{\citenamefont{Bang et~al.}(2002)\citenamefont{Bang, Krolikowski,
  Wyller, and Rasmussen}}]{Bang:02:pre}
\bibinfo{author}{\bibfnamefont{O.}~\bibnamefont{Bang}},
  \bibinfo{author}{\bibfnamefont{W.}~\bibnamefont{Krolikowski}},
  \bibinfo{author}{\bibfnamefont{J.}~\bibnamefont{Wyller}}, \bibnamefont{and}
  \bibinfo{author}{\bibfnamefont{J.~J.} \bibnamefont{Rasmussen}},
  \bibinfo{journal}{Phys. Rev. E} \textbf{\bibinfo{volume}{66}},
  \bibinfo{pages}{046619} (\bibinfo{year}{2002}).

\bibitem[{\citenamefont{Nath et~al.}(2007)\citenamefont{Nath, Pedri, and
  Santos}}]{Nath:07:pra}
\bibinfo{author}{\bibfnamefont{R.}~\bibnamefont{Nath}},
  \bibinfo{author}{\bibfnamefont{P.}~\bibnamefont{Pedri}}, \bibnamefont{and}
  \bibinfo{author}{\bibfnamefont{L.}~\bibnamefont{Santos}},
  \bibinfo{journal}{Phys. Rev. A} \textbf{\bibinfo{volume}{76}},
  \bibinfo{pages}{013606} (\bibinfo{year}{2007}).

\bibitem[{\citenamefont{Cuevas et~al.}(2009)\citenamefont{Cuevas, Malomed,
  Kevrekidis, and Frantzeskakis}}]{Cuevas:09:pra}
\bibinfo{author}{\bibfnamefont{J.}~\bibnamefont{Cuevas}},
  \bibinfo{author}{\bibfnamefont{B.~A.} \bibnamefont{Malomed}},
  \bibinfo{author}{\bibfnamefont{P.~G.} \bibnamefont{Kevrekidis}},
  \bibnamefont{and} \bibinfo{author}{\bibfnamefont{D.~J.}
  \bibnamefont{Frantzeskakis}}, \bibinfo{journal}{Phys. Rev. A}
  \textbf{\bibinfo{volume}{79}}, \bibinfo{pages}{053608}
  (\bibinfo{year}{2009}).

\bibitem[{\citenamefont{Lashkin}(2007)}]{Lashkin:07:pra}
\bibinfo{author}{\bibfnamefont{V.~M.} \bibnamefont{Lashkin}},
  \bibinfo{journal}{Phys. Rev. A} \textbf{\bibinfo{volume}{75}},
  \bibinfo{pages}{043607} (\bibinfo{year}{2007}).

\bibitem[{\citenamefont{Lashkin}(2008)}]{Lashkin:08:pra_a}
\bibinfo{author}{\bibfnamefont{V.~M.} \bibnamefont{Lashkin}},
  \bibinfo{journal}{Phys. Rev. A} \textbf{\bibinfo{volume}{78}},
  \bibinfo{pages}{033603} (\bibinfo{year}{2008}).

\bibitem[{\citenamefont{Zaliznyak and Yakimenko}(2008)}]{Zaliznyak:08:pla}
\bibinfo{author}{\bibfnamefont{Y.~A.} \bibnamefont{Zaliznyak}}
  \bibnamefont{and} \bibinfo{author}{\bibfnamefont{A.~I.}
  \bibnamefont{Yakimenko}}, \bibinfo{journal}{Phys. Lett. A}
  \textbf{\bibinfo{volume}{372}}, \bibinfo{pages}{2862} (\bibinfo{year}{2008}).

\bibitem[{\citenamefont{Lashkin et~al.}(2009)\citenamefont{Lashkin, Yakimenko,
  and Zaliznyak}}]{Lashkin:09:pscr}
\bibinfo{author}{\bibfnamefont{V.~M.} \bibnamefont{Lashkin}},
  \bibinfo{author}{\bibfnamefont{A.~L.} \bibnamefont{Yakimenko}},
  \bibnamefont{and} \bibinfo{author}{\bibfnamefont{Y.~A.}
  \bibnamefont{Zaliznyak}}, \bibinfo{journal}{Phys. Scr.}
  \textbf{\bibinfo{volume}{79}}, \bibinfo{pages}{035305}
  (\bibinfo{year}{2009}).

\bibitem[{\citenamefont{Nath et~al.}(2008)\citenamefont{Nath, Pedri, and
  Santos}}]{Nath:08}
\bibinfo{author}{\bibfnamefont{R.}~\bibnamefont{Nath}},
  \bibinfo{author}{\bibfnamefont{P.}~\bibnamefont{Pedri}}, \bibnamefont{and}
  \bibinfo{author}{\bibfnamefont{L.}~\bibnamefont{Santos}},
  \bibinfo{journal}{Phys. Rev. Lett.} \textbf{\bibinfo{volume}{101}},
  \bibinfo{pages}{210402} (\bibinfo{year}{2008}).

\bibitem[{\citenamefont{O'Dell et~al.}(2000)\citenamefont{O'Dell, Giovanazzi,
  Kurizki, and Akulin}}]{Odell:00}
\bibinfo{author}{\bibfnamefont{D.~S.} \bibnamefont{O'Dell}},
  \bibinfo{author}{\bibfnamefont{S.}~\bibnamefont{Giovanazzi}},
  \bibinfo{author}{\bibfnamefont{G.}~\bibnamefont{Kurizki}}, \bibnamefont{and}
  \bibinfo{author}{\bibfnamefont{V.~M.} \bibnamefont{Akulin}},
  \bibinfo{journal}{Phys. Rev. Lett.} \textbf{\bibinfo{volume}{84}},
  \bibinfo{pages}{5687} (\bibinfo{year}{2000}).

\bibitem[{\citenamefont{Turitsyn}(1985)}]{Turitsyn:85:tmf}
\bibinfo{author}{\bibfnamefont{S.~K.} \bibnamefont{Turitsyn}},
  \bibinfo{journal}{Theor, Mat. Fiz.} \textbf{\bibinfo{volume}{64}},
  \bibinfo{pages}{797} (\bibinfo{year}{1985}).

\bibitem[{\citenamefont{Giovanazzi et~al.}(2001)\citenamefont{Giovanazzi,
  O'Dell, and Kurizki}}]{Giovanazzi:01:pra}
\bibinfo{author}{\bibfnamefont{S.}~\bibnamefont{Giovanazzi}},
  \bibinfo{author}{\bibfnamefont{D.}~\bibnamefont{O'Dell}}, \bibnamefont{and}
  \bibinfo{author}{\bibfnamefont{G.}~\bibnamefont{Kurizki}},
  \bibinfo{journal}{Phys. Rev. A} \textbf{\bibinfo{volume}{63}},
  \bibinfo{pages}{031603(R)} (\bibinfo{year}{2001}).

\bibitem[{\citenamefont{Papadopoulos et~al.}(2007)\citenamefont{Papadopoulos,
  Wagner, Wunner, and Main}}]{Papadopoulos:07:pra}
\bibinfo{author}{\bibfnamefont{I.}~\bibnamefont{Papadopoulos}},
  \bibinfo{author}{\bibfnamefont{P.}~\bibnamefont{Wagner}},
  \bibinfo{author}{\bibfnamefont{G.}~\bibnamefont{Wunner}}, \bibnamefont{and}
  \bibinfo{author}{\bibfnamefont{J.}~\bibnamefont{Main}},
  \bibinfo{journal}{Phys. Rev. A} \textbf{\bibinfo{volume}{76}},
  \bibinfo{pages}{053604} (\bibinfo{year}{2007}).

\bibitem[{\citenamefont{Cartarius et~al.}(2008)\citenamefont{Cartarius, Fabcic,
  Main, and Wunner}}]{Cartarius:08:pra}
\bibinfo{author}{\bibfnamefont{H.}~\bibnamefont{Cartarius}},
  \bibinfo{author}{\bibfnamefont{T.}~\bibnamefont{Fabcic}},
  \bibinfo{author}{\bibfnamefont{J.}~\bibnamefont{Main}}, \bibnamefont{and}
  \bibinfo{author}{\bibfnamefont{G.}~\bibnamefont{Wunner}},
  \bibinfo{journal}{Phys. Rev. A} \textbf{\bibinfo{volume}{78}},
  \bibinfo{pages}{013615} (\bibinfo{year}{2008}).

\bibitem[{\citenamefont{Keles et~al.}(2008)\citenamefont{Keles, Sevincli, and
  Tanatar}}]{Keles:08:pra}
\bibinfo{author}{\bibfnamefont{A.}~\bibnamefont{Keles}},
  \bibinfo{author}{\bibfnamefont{S.}~\bibnamefont{Sevincli}}, \bibnamefont{and}
  \bibinfo{author}{\bibfnamefont{B.}~\bibnamefont{Tanatar}},
  \bibinfo{journal}{Phys. Rev. A} \textbf{\bibinfo{volume}{77}},
  \bibinfo{pages}{053604} (\bibinfo{year}{2008}).

\bibitem[{\citenamefont{Krolikowski et~al.}(2004)\citenamefont{Krolikowski,
  Nikolov, Neshev, Bang, Rasmussen, and Wyller}}]{Krolikowski:04:job}
\bibinfo{author}{\bibfnamefont{W.}~\bibnamefont{Krolikowski}},
  \bibinfo{author}{\bibfnamefont{N.}~\bibnamefont{Nikolov}},
  \bibinfo{author}{\bibfnamefont{D.}~\bibnamefont{Neshev}},
  \bibinfo{author}{\bibfnamefont{O.}~\bibnamefont{Bang}},
  \bibinfo{author}{\bibfnamefont{J.~J.} \bibnamefont{Rasmussen}},
  \bibnamefont{and} \bibinfo{author}{\bibfnamefont{J.}~\bibnamefont{Wyller}},
  \bibinfo{journal}{J. Opt. B.} \textbf{\bibinfo{volume}{6}},
  \bibinfo{pages}{288} (\bibinfo{year}{2004}).

\bibitem[{\citenamefont{Buccoliero
  et~al.}(2007{\natexlab{a}})\citenamefont{Buccoliero, Lopez-Aguayo, Skupin,
  Desyatnikov, Bang, Krolikowski, and Kivshar}}]{Buccoliero:07:PhysicaB}
\bibinfo{author}{\bibfnamefont{D.}~\bibnamefont{Buccoliero}},
  \bibinfo{author}{\bibfnamefont{S.}~\bibnamefont{Lopez-Aguayo}},
  \bibinfo{author}{\bibfnamefont{S.}~\bibnamefont{Skupin}},
  \bibinfo{author}{\bibfnamefont{A.}~\bibnamefont{Desyatnikov}},
  \bibinfo{author}{\bibfnamefont{O.}~\bibnamefont{Bang}},
  \bibinfo{author}{\bibfnamefont{W.}~\bibnamefont{Krolikowski}},
  \bibnamefont{and} \bibinfo{author}{\bibfnamefont{Y.~S.}
  \bibnamefont{Kivshar}}, \bibinfo{journal}{Physica B}
  \textbf{\bibinfo{volume}{394}}, \bibinfo{pages}{351}
  (\bibinfo{year}{2007}{\natexlab{a}}).

\bibitem[{\citenamefont{Skupin et~al.}(2008)\citenamefont{Skupin, Grech, and
  Krolikowski}}]{Skupin:08:oe}
\bibinfo{author}{\bibfnamefont{S.}~\bibnamefont{Skupin}},
  \bibinfo{author}{\bibfnamefont{M.}~\bibnamefont{Grech}}, \bibnamefont{and}
  \bibinfo{author}{\bibfnamefont{W.}~\bibnamefont{Krolikowski}},
  \bibinfo{journal}{Opt. Express} \textbf{\bibinfo{volume}{16}},
  \bibinfo{pages}{9118} (\bibinfo{year}{2008}).

\bibitem[{\citenamefont{Buccoliero
  et~al.}(2007{\natexlab{b}})\citenamefont{Buccoliero, Desyatnikov,
  Krolikowski, and Kivshar}}]{Buccoliero:07}
\bibinfo{author}{\bibfnamefont{D.}~\bibnamefont{Buccoliero}},
  \bibinfo{author}{\bibfnamefont{A.~S.} \bibnamefont{Desyatnikov}},
  \bibinfo{author}{\bibfnamefont{W.}~\bibnamefont{Krolikowski}},
  \bibnamefont{and} \bibinfo{author}{\bibfnamefont{Y.~S.}
  \bibnamefont{Kivshar}}, \bibinfo{journal}{Phys. Rev. Lett.}
  \textbf{\bibinfo{volume}{98}}, \bibinfo{pages}{053901}
  (\bibinfo{year}{2007}{\natexlab{b}}).

\bibitem[{\citenamefont{Ginibre and Velo}(1980)}]{ginibre:80:MZ}
\bibinfo{author}{\bibfnamefont{J.}~\bibnamefont{Ginibre}} \bibnamefont{and}
  \bibinfo{author}{\bibfnamefont{G.}~\bibnamefont{Velo}},
  \bibinfo{journal}{Mathematische Zeitschrift} \textbf{\bibinfo{volume}{170}},
  \bibinfo{pages}{109} (\bibinfo{year}{1980}).

\bibitem[{\citenamefont{Fr{\"o}hlich et~al.}(2002)\citenamefont{Fr{\"o}hlich,
  Tsai, and Yau}}]{Froelich:02:CMP}
\bibinfo{author}{\bibfnamefont{J.}~\bibnamefont{Fr{\"o}hlich}},
  \bibinfo{author}{\bibfnamefont{T.}~\bibnamefont{Tsai}}, \bibnamefont{and}
  \bibinfo{author}{\bibfnamefont{H.}~\bibnamefont{Yau}},
  \bibinfo{journal}{Communications in Mathematical Physics}
  \textbf{\bibinfo{volume}{225}}, \bibinfo{pages}{223} (\bibinfo{year}{2002}).

\bibitem[{\citenamefont{Skryabin et~al.}(2002)\citenamefont{Skryabin, McSloy,
  and Firth}}]{Skryabin:07:pre}
\bibinfo{author}{\bibfnamefont{D.~V.} \bibnamefont{Skryabin}},
  \bibinfo{author}{\bibfnamefont{J.~M.} \bibnamefont{McSloy}},
  \bibnamefont{and} \bibinfo{author}{\bibfnamefont{W.~J.} \bibnamefont{Firth}},
  \bibinfo{journal}{Phys. Rev. E} \textbf{\bibinfo{volume}{66}},
  \bibinfo{pages}{055602(R)} (\bibinfo{year}{2002}).

\bibitem[{\citenamefont{Desyatnikov et~al.}(2005)\citenamefont{Desyatnikov,
  Sukhorukov, and Kivshar}}]{Desyatnikov:05}
\bibinfo{author}{\bibfnamefont{A.~S.} \bibnamefont{Desyatnikov}},
  \bibinfo{author}{\bibfnamefont{A.~A.} \bibnamefont{Sukhorukov}},
  \bibnamefont{and} \bibinfo{author}{\bibfnamefont{Y.~S.}
  \bibnamefont{Kivshar}}, \bibinfo{journal}{Phys. Rev. Lett.}
  \textbf{\bibinfo{volume}{95}}, \bibinfo{pages}{203904}
  (\bibinfo{year}{2005}).

\bibitem[{\citenamefont{Rozanov}(2004)}]{Rosanov:os:96:405}
\bibinfo{author}{\bibfnamefont{N.~N.} \bibnamefont{Rozanov}},
  \bibinfo{journal}{Opt Spectrosc} \textbf{\bibinfo{volume}{96}},
  \bibinfo{pages}{405} (\bibinfo{year}{2004}).

\bibitem[{\citenamefont{Malomed}(2002)}]{variational}
\bibinfo{author}{\bibfnamefont{B.~A.} \bibnamefont{Malomed}},
  \bibinfo{journal}{Prog. Opt.} \textbf{\bibinfo{volume}{43}},
  \bibinfo{pages}{71} (\bibinfo{year}{2002}).

\bibitem[{\citenamefont{Skupin et~al.}(2006)\citenamefont{Skupin, Bang,
  Edmundson, and Krolikowski}}]{Skupin:pre:73:066603}
\bibinfo{author}{\bibfnamefont{S.}~\bibnamefont{Skupin}},
  \bibinfo{author}{\bibfnamefont{O.}~\bibnamefont{Bang}},
  \bibinfo{author}{\bibfnamefont{D.}~\bibnamefont{Edmundson}},
  \bibnamefont{and}
  \bibinfo{author}{\bibfnamefont{W.}~\bibnamefont{Krolikowski}},
  \bibinfo{journal}{Phys. Rev. E} \textbf{\bibinfo{volume}{73}},
  \bibinfo{pages}{066603} (\bibinfo{year}{2006}).

\bibitem[{\citenamefont{Lopez-Aguayo et~al.}(2006)\citenamefont{Lopez-Aguayo,
  Desyatnikov, Kivshar, Skupin, Krolikowski, and Bang}}]{Lopez:ol:31:1100}
\bibinfo{author}{\bibfnamefont{S.}~\bibnamefont{Lopez-Aguayo}},
  \bibinfo{author}{\bibfnamefont{A.}~\bibnamefont{Desyatnikov}},
  \bibinfo{author}{\bibfnamefont{Y.~S.} \bibnamefont{Kivshar}},
  \bibinfo{author}{\bibfnamefont{S.}~\bibnamefont{Skupin}},
  \bibinfo{author}{\bibfnamefont{W.}~\bibnamefont{Krolikowski}},
  \bibnamefont{and} \bibinfo{author}{\bibfnamefont{O.}~\bibnamefont{Bang}},
  \bibinfo{journal}{Opt. Lett} \textbf{\bibinfo{volume}{31}},
  \bibinfo{pages}{1100} (\bibinfo{year}{2006}).

\end{thebibliography}

\end{document}